\documentclass[sn-mathphys-num]{sn-jnl}

\usepackage{graphicx}
\usepackage{multirow}
\usepackage{amsmath,amssymb,amsfonts}
\usepackage{amsthm}
\usepackage{mathrsfs}
\usepackage[title]{appendix}
\usepackage{xcolor}
\usepackage{textcomp}
\usepackage{manyfoot}
\usepackage{booktabs}
\usepackage{algorithm}
\usepackage{algorithmicx}
\usepackage{algpseudocode}
\usepackage{listings}
\usepackage{hyperref}
\usepackage{braket}
\usepackage{lineno} 
\usepackage{caption}

\theoremstyle{thmstyleone}

\theoremstyle{thmstyletwo}%

\theoremstyle{thmstylethree}%

\begin{document}


\title[Article Title]{Fiber transmission of cluster states via multi-level time-bin encoding}




\author[1,2]{\fnm{Philip} \sur{Rübeling}}
\equalcont{These authors contributed equally to this work.}

\author[1,2]{\fnm{Robert} \sur{Johanning}}
\equalcont{These authors contributed equally to this work.}

\author[1,2]{\fnm{Jan} \sur{Heine}}

\author[1,2]{\fnm{Oleksandr V.} \sur{Marchukov}}

\author*[1,2]{\fnm{Michael} \sur{Kues}}\email{michael.kues@iop.uni-hannover.de}

\affil[1]{\orgdiv{Institute of Photonics (IOP)}, \orgname{Leibniz University Hannover}, \orgaddress{\street{Nienburger Straße 17}, \city{Hannover}, \postcode{30167}, \state{Lower Saxony}, \country{Germany}}}

\affil[2]{\orgdiv{Cluster of Excellence PhoenixD}, \orgname{Leibniz University Hannover}, \orgaddress{\street{Am Welfengarten 1}, \city{Hannover}, \postcode{30167}, \state{Lower Saxony}, \country{Germany}}}

\setlength{\parindent}{0pt}

\abstract{The next generation of telecommunication networks will rely on the transmission of complex quantum states to enable secure and transformative information processing, utilizing entanglement and superposition. Cluster states – multipartite entangled states that retain entanglement under local measurements – are a vital resource for quantum networking applications such as blind photonic quantum computing, quantum state teleportation and all-photonic quantum repeaters. However, the transmission of cluster states over optical fiber has remained elusive with previous approaches. Here, we demonstrate the first transmission of a four-qubit cluster state over 25 km of single-mode fiber by using a two-photon multi-level time-bin encoding. We directly generate the state by exploiting coherent control of a parametric generation process, rendering a resource-intensive controlled-phase gate obsolete. To enable efficient and reconfigurable projective measurements on the multi-level time-bin encoded state, we introduce chirped-pulse modulation and implement the first time-bin beam splitter, allowing us to certify genuine multipartite entanglement and to demonstrate one-way computing operations. Our approach enables the transmission of complex quantum states over long-distance fibers, permitting the implementation of multipartite protocols and laying the foundation for large-scale quantum resource networks.}


\maketitle


\section{Introduction}\label{sec1}

Photonic quantum technologies~\cite{Obrien_2009} utilize superposition and entanglement of light to drive the next revolution in information technology, promising unprecedented possibilities in computing~\cite{Maring_2024} and secure communications~\cite{Gisin_2007}. The backbone of these technologies are quantum networks~\cite{Wehner_2018, Azuma_2023} interconnecting remote users to transfer quantum states as a resource of entanglement~\cite{Wen_2022, Alshowkan_2021}.\\

Notably, the existing telecommunication infrastructure can be harnessed to implement quantum networks via optical single-mode fiber which paves the way to robust and high-speed quantum state transfer. At present, deployed fiber networks are being used to distribute bipartite entanglement over hundreds of kilometers~\cite{Neumann_2022} enabling applications such as quantum key distribution~\cite{Honjo_2008} and distributed quantum computing~\cite{vanLeent_2022, Main_2025}.\\

In recent years, multipartite entanglement in quantum networks has gained attention, revealing unparalleled capabilities for secure communication and anonymous data processing, exceeding the frontiers of bipartite entanglement~\cite{Pan_2012}. In particular, cluster states — multipartite quantum states that remain entangled when measured locally — provide a vital resource for advanced quantum networking applications such as blind photonic quantum computing~\cite{Barz_2012, Broadbent_2009}, quantum state teleportation~\cite{Da_2007} and all-photonic quantum repeaters~\cite{Azuma_2015}.\\

However, a critical bottleneck to quantum networking is the transmission of cluster states over optical single-mode fiber~\cite{vanDam_2024}. Specifically, quantum state encodings with one photon per qubit — such as polarization-encoded states~\cite{Proietti_2021} — suffer from optical losses exponentially growing with the number of photons. In addition, spatial encodings for multipartite entanglement such as orbital angular momentum~\cite{Lib_2024} are inherently incompatible with single-mode fibers and thus cannot be transmitted over existing telecommunications infrastructure.\\

In contrast, encodings in the time and frequency domain~\cite{Lu_2023} are natively compatible with optical fibers and today’s telecommunications infrastructure~\cite{Ruebeling_2024, Kashi_2025} and can be integrated on a photonic chip~\cite{Reimer_2016, Mahmudlu_2023, Finco_2024}. While cluster states have already been generated in the time and frequency domain, these implementations rely on resource-intensive experimental setups~\cite{Larsen_2019} and active stabilization in multiple degrees of freedom~\cite{Reimer_2019}, rendering their transmission beyond reach. To achieve transmission of cluster states, more robust approaches are required, retaining the high information capacity of the time and frequency domain.\\

Here, we demonstrate the transmission of a four-qubit cluster state over 25 kilometers of optical single-mode fiber; fully compatible with the existing telecommunications infrastructure. We encoded the cluster state into multi-level time-bin entangled photons pairs and directly generated this state without a controlled phase gate, massively reducing the experimental overhead. Importantly, we demonstrate one-way quantum operations on the transmitted cluster state that lay the foundation for advanced quantum network applications, including blind photonic quantum computing and all-photonic quantum repeaters. To achieve this, we developed a reconfigurable quantum pulse shaping technique, based on highly-dispersive chirped fiber Bragg gratings and electro-optic phase modulation to implement a tunable time-bin beam splitter.

\section{Results}\label{sec2}

In our experiment, a four-qubit cluster states was transmitted over 25 km of optical fiber, see Fig. \ref{fig:Figure_01} (a). To this end, we generated time-bin entangled photon pairs by cascading second harmonic generation and spontaneous parametric down-conversion excited with four coherent pulses at telecommunications wavelength, see Fig. \ref{fig:Figure_01} (b). We encoded the four-qubit cluster state into four time-bins of the signal and idler photon, respectively. Each photon encodes two qubits:  the basis states of the first qubit (called T-level qubit) are defined by grouping the first two and the last two time-bins, while for the second qubit (called t-level qubit), the basis states are defined by grouping the first and third time-bin and the second and fourth time-bin. Following this encoding, the four-qubit cluster state can be written as

\begin{equation}
    |\Psi_C \rangle =  \frac{1}{2} \big( 
    |0_{T,s} 0_{T,i} 0_{t,s} 0_{t,i} \rangle +
    |0_{T,s} 0_{T,i} 1_{t,s} 1_{t,i} \rangle +
    |1_{T,s} 1_{T,i} 0_{t,s} 0_{t,i} \rangle -
    |1_{T,s} 1_{T,i} 1_{t,s} 1_{t,i} \rangle 
    \big),
\end{equation}

where each basis state is labeled by signal or idler photon (\(s, i\)) and with respect to the two levels (\(T, t\)). In previous work, a controlled-phase gate was used to transform hyper-entangled states into cluster states~\cite{Kiesel_2005, Reimer_2019}, introducing optical loss to the quantum state and adding experimental complexity. In contrast, in our encoding scheme, a gate is not required. Instead, we utilized an electro-optic phase modulator (EOPM) to apply a phase of \(\pi/2\) to the fourth excitation pulse, see Fig. \ref{fig:Figure_01} (b), to directly generate a cluster state via the cascaded parametric process (see Methods). This approach shifts the manipulation of the photon pairs to the coherent excitation pulses, reducing optical loss for the quantum state and experimental complexity – critical aspects for quantum network deployments.\\

After the generation, we transmitted the cluster states over 25 km of single-mode optical fiber followed by a dispersion compensation module to emulate a fiber link within a network.  Thermal drifts of the fiber length introduce variations in photon arrival time, which can render the detected time-bins indistinct, especially for long integration periods. To compensate for this, we actively stabilized the fiber length with a motorized optical delay line, by applying a feedback from the time-of-flight of the transmitted photons.\\

We certified the presence of genuine multipartite entanglement of the transmitted cluster state. To this end, we employ an entanglement witness~\cite{Lewenstein_2000} - derived from the stabilizer group of the respective state - which is an established method to verify multipartite entanglement in complex quantum states (see Methods). To compute a witness for the cluster state, local measurements on each qubit need to be performed in two mutually unbiased bases~\cite{Kiesel_2005}. In our case, this translates to projective measurements in the Z-basis and in the X-basis that need to be performed on the t-level qubits and the T-level qubits. \\

Typically, projective measurements on time-bin encoded qubits are implemented by imbalanced fiber interferometers~\cite{Roztocki_2021}. However, these interferometers have a fixed imbalance to match a single time-bin spacing, limiting their reconfigurability. In addition, for our multi-level time-bin encoding, four imbalanced fiber interferometers, each stabilized to a reference laser would be required (for each level for signal and idler, respectively) to obtain the required projective measurement, adding significant experimental complexity. Moreover, fiber interferometers are limited to one superposition projection per measurement setting.\\

To overcome these technological limitations, we introduce chirped pulse modulation (CPM); a novel technique for robust and reconfigurable processing of ultrafast time-bin quantum states, extending spectral pulse shaping for coherent light~\cite{Weiner_1988}. Here, our experimental setup includes an electro-optic phase modulator (EOPM) located in-between two highly dispersive chirped fiber Bragg gratings (CFBGs) of opposite dispersion. As shown in Fig. \ref{fig:Figure_01} (c), CPM processes time-bins in three sequential steps: (i) The CFBG with positive chirp ($+10$ ns/nm) maps the spectrum of the time-bins to the time domain, approximating a dispersive Fourier transform~\cite{Godin_2022, Widomski_2024} (see Methods). (ii) The spectral phase is modulated by an EOPM on which a radio-frequency waveform is applied. (iii) The CFBG with negative chirp ($-10$ ns/nm) maps the spectral components back into the time domain, approximating a inverse dispersive Fourier transform.  As a result, coherent replica of the time-bins are generated at integer multiples of \(\Delta t = |\beta|_2 \Omega\), where \(\beta_2\) is the group velocity dispersion of the CFBGs and \(\Omega\) is the angular frequency of the radio-frequency waveform (see Methods). \\

To perform CPM on our cluster state, we drove the EOPM with an arbitrary waveform generator (AWG) at radio-frequencies of 1.25 GHz and 3.75 GHz to superimpose time-bins of 100 ps (t-level qubits) and 300 ps (T-Level qubits) spacing, respectively. To ensure phase synchronization across the setup, a single AWG was used to drive the modulators for the excitation pulse shaping and the CPM. Importantly, we controlled the phase of the radio-frequency signal with respect to the arrival time of the time-bins and adjusted the signal amplitude to implement a reconfigurable time-bin beam splitter, see Fig. \ref{fig:Figure_01} (c). We applied a single segmented waveform at the AWG, see Fig. \ref{fig:Figure_02} (a), realizing nine different beam splitter settings, see Fig. \ref{fig:Figure_02} (b). The waveform was repeated every 180 ns, allowing all settings being measured in parallel, making our approach robust to fluctuations in the count rates, caused e.g. by polarization rotations.\\

Subsequently, we used a programmable filter to select a 15 GHz bandwidth around 193.7 THz for the signal and 193.1 THz for the idler photons, and routed them to superconducting nanowire single-photon detectors. This filtering ensured that the photons remained reasonably within their single-mode bandwidth of 11.9 GHz. Further, the frequency offset of 600 GHz between signal and idler photons resulted in a temporal separation of 48 ns after the first CFBG. This enabled independent phase modulation of signal and idler photons by the EOPM. \\

From the nine configurations of the time-bin beam splitter, we extracted 48 projective measurements, see Fig. \ref{fig:Figure_03} (a), which we used to compute the stabilizers of the cluster state, see Fig. \ref{fig:Figure_03} (b). From the stabilizers, we calculated the witness \(W_C=(-0.80 \pm 0.04)\), verifying the presence of multipartite entanglement after 25 km of optical fiber with a confidence of 20 standard deviations, see Fig. \ref{fig:Figure_03} (c).\\

To showcase the applicability of the transmitted cluster state for quantum networking applications, we performed one-way computing operations~\cite{Raussendorf_2001}. To accomplish this, we measured two-qubit quantum interference on the cluster state by projecting the t-level qubits into one of the Z-basis  states, while interfering the T-level qubits and vice versa, see Fig. \ref{fig:Figure_03} (d-g). The visibilities of all measurements exceed $92.2\%$, clearly violating the CHSH inequality~\cite{Collins_2002} proving the residual bipartite entanglement after partial projective measurements on the transmitted cluster state.

\section{Discussion}\label{sec3}

In this work, we have demonstrated the transmission of a cluster state over optical single-mode fiber, fully compatible with today's telecommunications infrastructure. We verified that the transmitted cluster state is applicable for one-way computing, proving its utility for quantum networking. For this achievement, we generated the cluster state in a multi-level time-bin encoding. In contrast to previous time-frequency cluster state implementations~\cite{Reimer_2019}, our scheme does not require a controlled phase gate, significantly reducing the experimental complexity. \\

To manipulate the cluster state, we introduced a tunable time-bin beam splitter implemented by the chirped pulse modulation technique. In contrast to fiber interferometers, which are limited to one fixed time-bin spacing, our technique is fully reconfigurable and can superimpose ultrafast time-bins on multiple time-scales. \\

To extend the number of qubits in our time-bin encoded cluster state, additional levels of time-bins are required, see extended data Fig.~\ref{fig:Extended_Data_Figure_02},  adding two qubits each (one for signal, one for idler). While we can generate these states straightforwardly by leveraging arbitrary phase control over the excitation pulses, currently our system is limited by the dispersion of the CFBGs, diminishing the interference visibility to \(< 80\%\) for time-bins separated by more than 500 ps (see Methods). This can be overcome by cascading multiple CFBGs~\cite{Lukens_2018}, for example providing dispersion of 150 ns/nm~\cite{Proximion_2025}, enabling us to control five levels of time-bins, corresponding to a 10-qubit cluster state.\\

While we used a single frequency channels of 15 GHz bandwidth for signal and idler, respectively, our approach supports high-capacity processing of complex quantum states by leveraging spectral multiplexing (see Methods). This requires independent phase modulation of each frequency channel, which can be implemented, for example, leveraging arrayed waveguide gratings~\cite{Wang_2024} and chip-integrated electro-optic phase modulators~\cite{Wang_2018}. Remarkably, the control electronics require signal speeds of only a few gigahertz, which can be addressed using standard integrated electronic circuits~\cite{Atabaki_2018}. To deploy our approach in a real-world fiber network, sending and receiving nodes need to be synchronized temporally. This can be achieved by radio-frequency over fiber technology~\cite{Lim_2020}, which was recently demonstrated in a fiber-based quantum network~\cite{Chapman_2025}. \\

More broadly, our results are the first step towards the long-distance transmission of cluster states in quantum networks, paving the way for the next generation of quantum networks with applications in blind photonic quantum computing~\cite{Wei_2025}, all-photonic quantum repeaters~\cite{Azuma_2015}, cluster-state quantum teleportation~\cite{Da_2007} and multipartite quantum communications~\cite{Sun_2016, Rueckle_2023}. We anticipate that our demonstration of complex quantum state transfer will stimulate new directions at the intersection of quantum information science, distributed computing and artificial intelligence such as quantum-certified data integrity in cloud-based applications~\cite{Li_2024, Eisert_2020}.


\begin{figure}[p]
    \includegraphics[width=1.0\textwidth]{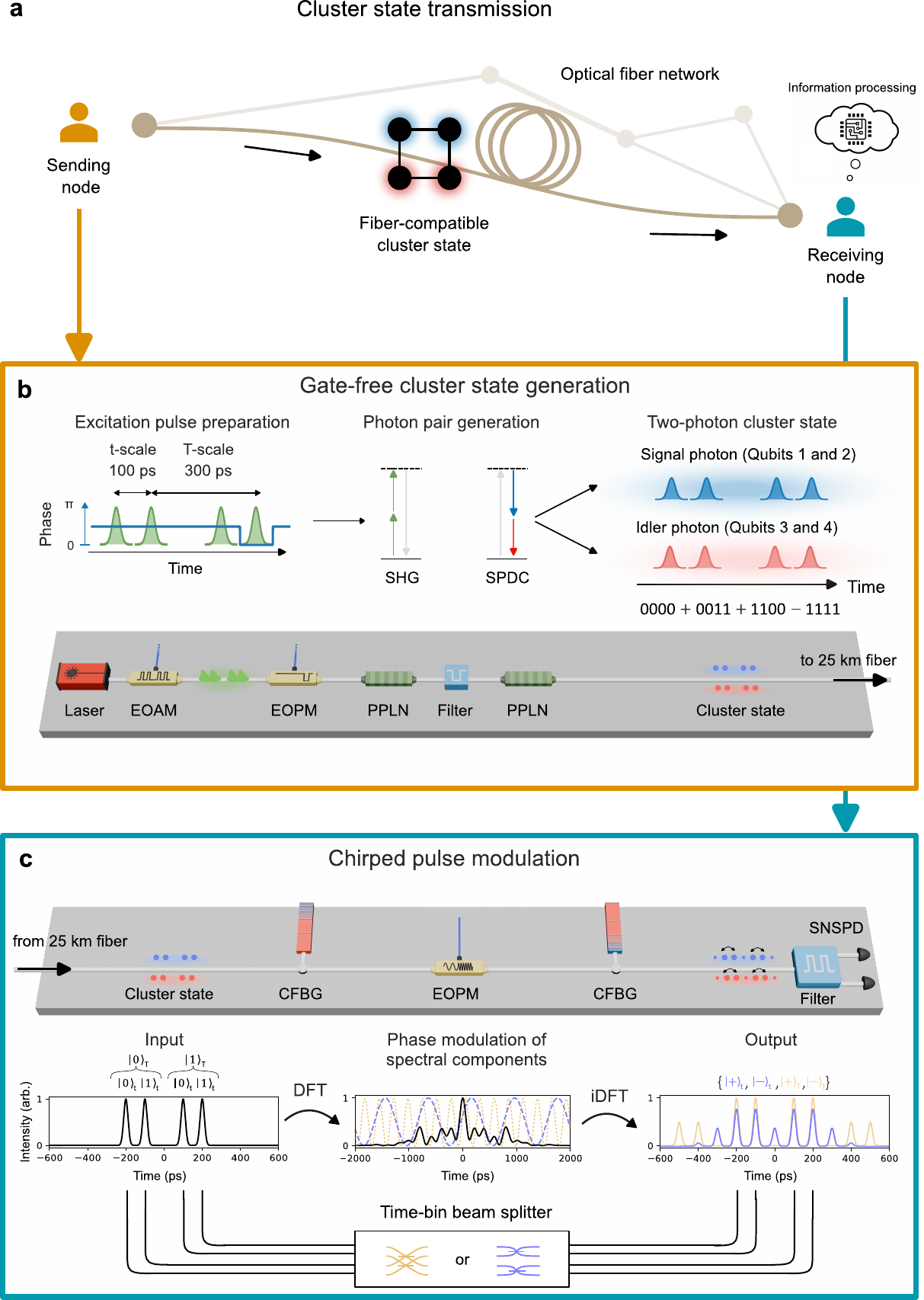}
    \centering
    \captionof{figure}{\textbf{Schematic of the four-qubit cluster state transmission experiment.} (a) A pair of time-bin entangled photons encoded the four-qubit cluster state which was transmitted over 25 kilometers of fiber, emulating a link in a fiber optical network.}
    \label{fig:Figure_01}
\end{figure}

\pagebreak

\begin{nolinenumbers}
\textbf{Fig.~\ref{fig:Figure_01} (continued).} Two qubits of the cluster state were encoded into the signal and idler photon (red and blue shading), respectively. The sending node prepared the cluster states and transmitted it to the receiving node, enabling distributed quantum information processing. (b) To implement gate-free cluster state generation, light from a continuous-wave laser was sent through an electro-optic amplitude modulator (EOAM) to generate four coherent excitation pulses (green). An electro-optic phase modulator (EOPM) applied a relative phase of \(\pi/2\) to the fourth of the excitation pulses, which were upconverted via second harmonic generation (SHG) in a periodically poled lithium niobate waveguide (PPLN). A filter suppressed the fundamental and spontaneous parametric downconversion (SPDC) in a PPLN generated the signal (blue) and idler photons (red) forming the two-photon four-qubit cluster state. (c) To implement chirped pulse modulation after the transmission of the cluster state, two chirped fiber Bragg gratings (CFBG) of opposite dispersion and an EOPM were used. The mechanism of chirped pulse modulation is shown for four input pulses (left). Propagation through the CFBG with positive dispersion implements an approximate dispersive Fourier transform (DFT), i.e., the frequency components of the input pulses are mapped into the time domain (center). Here, the EOPM modulates the dispersed spectral components. The dashed lines indicate the radio-frequency waveform that is applied to the EOPM; modulation at 3.75 GHz is depicted in orange, mixing time-bins of the T-scale and modulation at 1.25 GHz is depicted in purple, mixing time-bins of the t-scale. Propagation through the negatively dispersive CFBG implements an approximate inverse dispersive Fourier transform (iDFT) recompressing the spectral components in the time domain. The output time-bins (right) are superposition states of the input time-bins. The photons were routed to a programmable filter and detected by two superconducting nanowire single photon detectors (SNSPD). As shown at the bottom, this configuration implements a tunable time-bin beam splitter, either superimposing time-bins on the T-scale (orange) or on the t-scale (purple).\\
\end{nolinenumbers}

\begin{figure}[p]
    \includegraphics[width=1.0\textwidth]{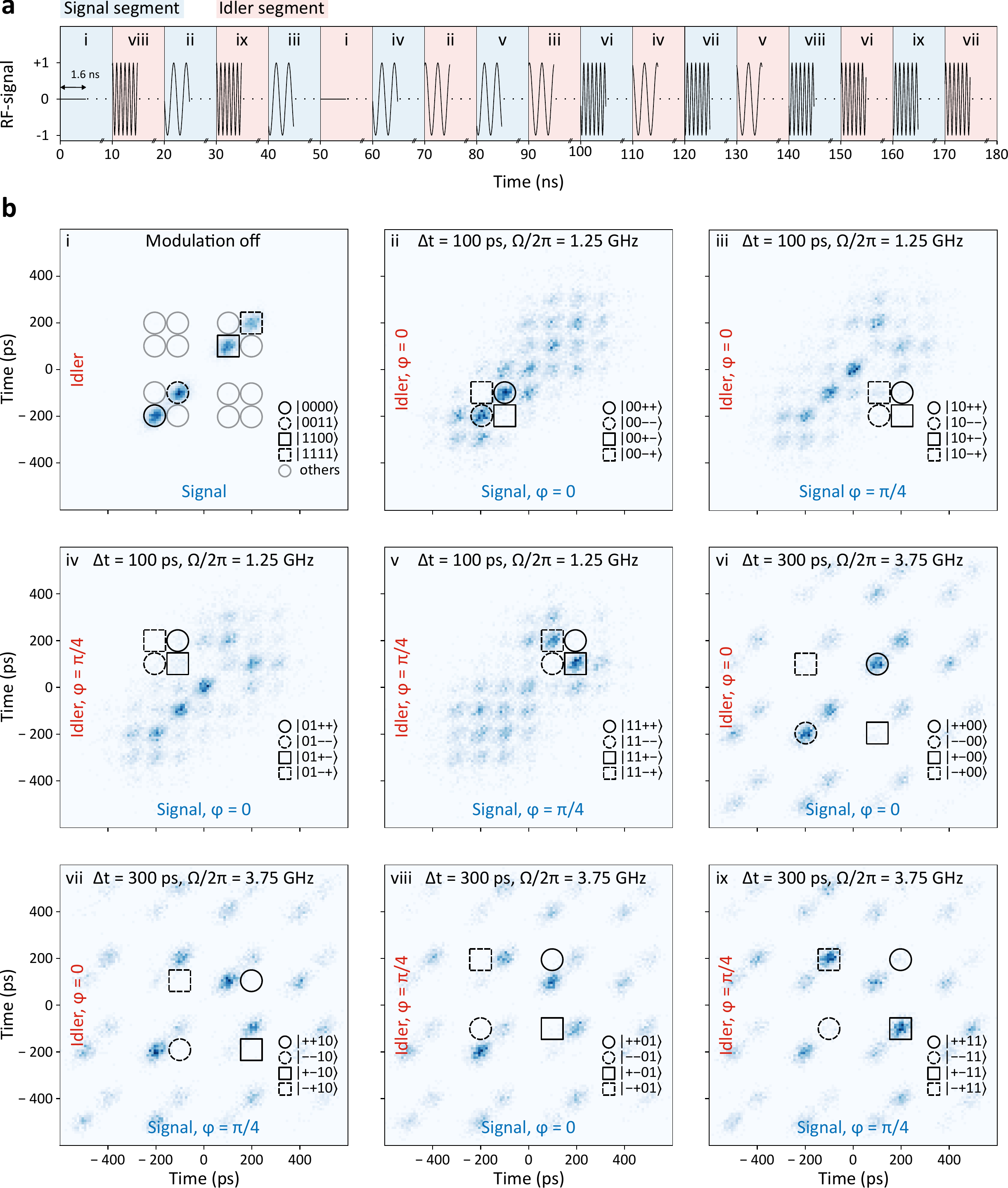}
    \centering
    \captionof{figure}{\textbf{Radio-frequency waveform used for chirped pulse modulation and corresponding joint temporal intensity measurements for the cluster state.} (a) The radio-frequency waveform applied to the chirped pulse modulation was composed of 18 segments each 10 ns long (here, only the first 1.6 ns are displayed for each segment). Signal and idler photons were temporally separated by \~ 48 ns after the first chirped fiber Bragg grating, enabling their independent modulation. Red and blue colors indicate whether signal or idler photons were modulated by the respective segment.}
    \label{fig:Figure_02}
\end{figure}

\pagebreak

\begin{nolinenumbers}
\textbf{Fig.~\ref{fig:Figure_02} (continued).} Two modulation tones (1.25 GHz for modulation at 100 ps on the t-scale and 3.75 GHz for modulation at 300 ps on the T-scale) with different phase relative to the time-bins were used to access all relevant 48 projections necessary to construct the witness operator. (b) Each pair of complementary waveform segments (labeled with Latin letters) generates one out of the nine displayed joint temporal intensity patterns. All 48 projections required for witness reconstruction can be derived from these measurements. Only a subset of projections was takes for each measurement setting, which are indicated by the circles and boxes. The projections in the other bins were not orthogonal.
\end{nolinenumbers}

\begin{figure}[p]
    \includegraphics[width=1.0\textwidth]{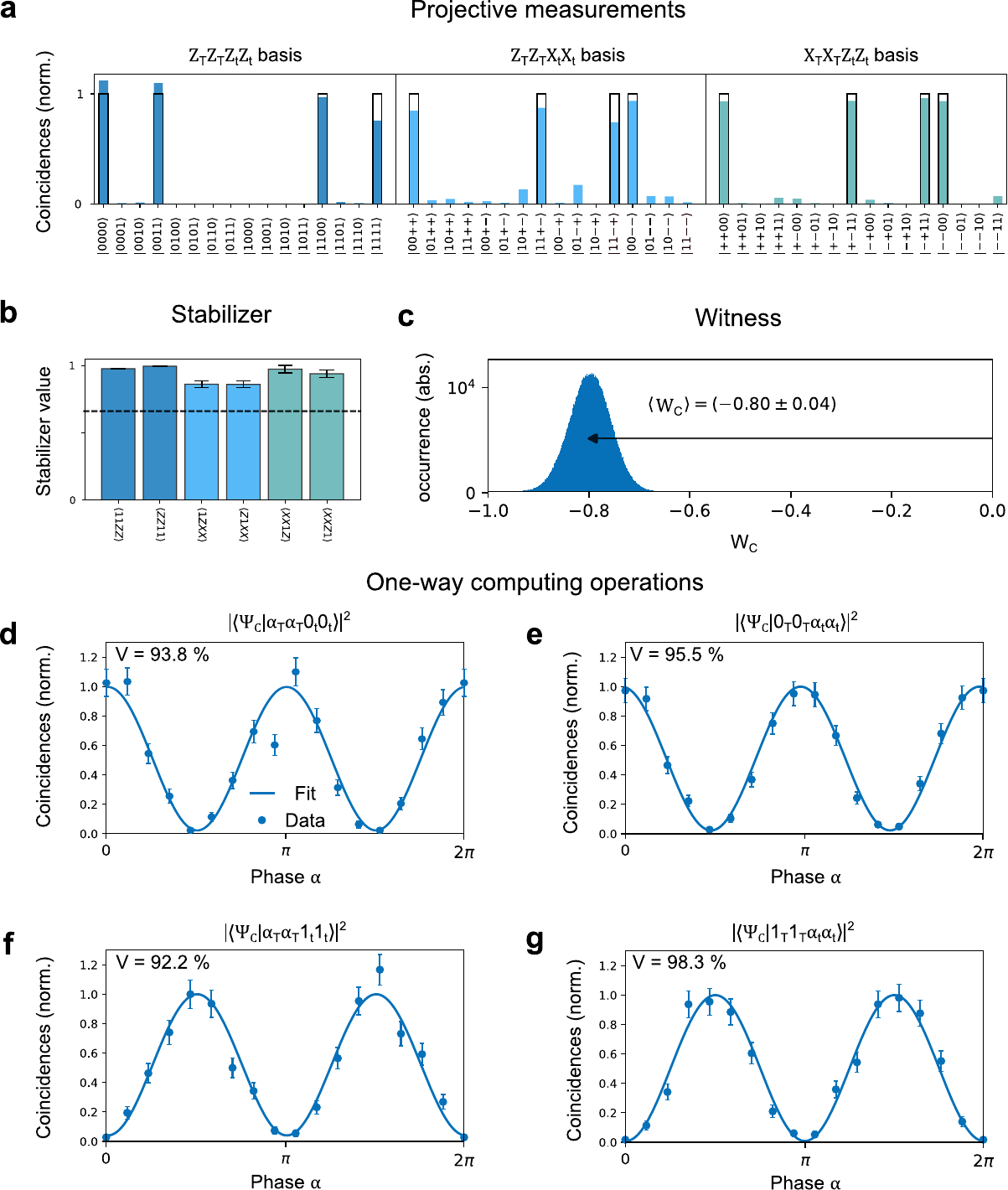}
    \centering
    \captionof{figure}{\textbf{Cluster state witness and one-way computing operations after transmission over 25 km of fiber.} Cluster state witness and one-way computing operations after transmission over 25 km of fiber. (a) 48 Projective measurements corresponding to three different bases, i.e. $Z_{T}Z_{T}Z_{t}Z_{t}$-basis (measurements in the computational basis), $Z_{T}Z_{T}X_{t}X_{t}$-basis (superpositions of the t-scale qubits) and $X_{T}X_{T}Z_{t}Z_{t}$-basis (superpositions of the T-scale qubits). Noise contributions are more dominant in the $Z_{T}Z_{T}X_{t}X_{t}$-basis due to linear crosstalk and the jitter of the single photon detectors (17 ps). The projections are normalized (see Methods) and the black boxes indicate expected values for the ideal cluster state.} 
    \label{fig:Figure_03}
\end{figure}

\pagebreak

\begin{nolinenumbers}
\textbf{Fig.~\ref{fig:Figure_03} (continued).} (b) Stabilizer values of the four-qubit cluster state. The black dashed line indicates the (mean) threshold value of 2/3 that all stabilizers need to exceed to verify multipartite entanglement. (c) The witness $W_{C} = (-0.80 \pm 0.04)$ is obtained from the stabilizers. The uncertainty of 20 standard deviations was estimated by a Monte-Carlo simulation. The histogram shows the (Gaussian) distribution of the witness values for $10^{6}$ samples of the Monte-Carlo simulation (see Methods). (d-g) To prove one-way quantum computing capabilities (e.g. gate teleportation), we measured quantum interference on the transmitted cluster state. Here, two qubits were measured in a basis rotated in the XY-plane (with phase $\alpha$) and the two remaining qubits were measured in the Z-basis (see Methods). By fitting the raw data (no background subtraction) to the expected sinusoidal curves, we achieved visibilities of (d) 93.8 \% for the projection onto $|\alpha_{T,s}, \alpha_{T,i}, 0_{t,s}, 0_{t,i} \rangle$, (e) 95.5 \% for the projection onto $|0_{T,s},0_{T,i},\alpha_{t,s}, \alpha_{t,i} \rangle$, (f) 92.2 \% for the projection onto $|\alpha_{T,s}, \alpha_{T,i}, 1_{t,s}, 1_{t,i} \rangle$ and (g) 98.3 \% for the projection onto $|1_{T,s},1_{T,i},\alpha_{t,s}, \alpha_{t,i} \rangle$, all above the CHSH-threshold of 70.7 \%.
\end{nolinenumbers}

\newpage

\section{Methods}

\subsection{Experimental setup}

To prepare coherent excitation pulses, we used a continuous-wave laser (Koheras Adjustik HP15, NKT Photonics) centered at 193.4 THz. Its output field was modulated with an electro-optic amplitude modulator (IM-1550-40-V-PM, Optilab) driven by an arbitrary waveform generator (M8194A, 120 GSa/s, Keysight). The amplitude modulator was used to shape a train of four excitation pulses, where each pulse had a duration of 37 ps (full-width at half maximum). We used an irregular temporal spacing for the pulses; 100 ps between the first and second pulse, 200 ps between the second and third pulse, and 100 ps between the third and fourth pulse. The entire pulse sequence was repeated every 20 ns, corresponding to 50 MHz repetition rate. To ensure stable operation of the amplitude modulator, we split 50\% of the signal to a bias controller (ABC, ID Photonics). \\

After creating the pulse pattern, we applied a phase of \(0\) to the first three pulses and a phase of \(\pi/2\) to the fourth pulse with an electro-optic phase modulator (EOSpace Inc.), which was driven by the same arbitrary waveform generator. Next, we used an erbium-doped fiber amplifier (Lumibird Inc.) to increase the average optical power to 5 mW. We launched this pulse train into a periodically-poled lithium niobate waveguide (PPLN, Covesion Ltd.) with 40 GHz phase-matching bandwidth to perform second-harmonic generation (SHG), converting the excitation pulses at 193.4 THz (1550 nm) to 386.8 THz (775 nm). In this process, two fundamental photons were converted into a single photon at the second harmonic frequency. For the phase-shifted fourth excitation pulse, this process doubles the previous applied phase to \(\pi\). Subsequently, we filtered the fundamental (193.4 THz) and passed the 775 nm light through a PPLN to directly generate the two-photon cluster state via spontaneous parametric down-conversion (SPDC). \\

The created cluster state was transmitted over 25 km of commercial low-loss single-mode fiber (SMF-28 Ultra, Corning Inc.) with 5.3 dB insertion loss. The fiber was wound on a spool and thermally isolated in a polystyrene box, experiencing temperature fluctuations of $< 0.1$ K. Assuming a thermal sensitivity of 36.8 ps/(K·km)~\cite{Sliwczynski_2010}, the resulting time-of-flight jitter is estimated to be 92 ps. We reduced this jitter to 1.7 ps by active stabilization of the fiber length. Specifically, we monitored the single-photon counts in intervals of 15 minutes to find the time-of-flight deviation and corrected for it with a motorized delay line (MDL-02, Luna Innovations),which was placed in the path of the excitation pulses to mitigate losses in the cluster states. The dispersion of the single-mode fiber was compensated with a dispersion compensating module (insertion loss 2.4 dB, dispersion of $-450$ ps/nm). \\

To implement chirped pulse modulation in our experiment, we used CFBGs with a positive dispersion of +9988 ps/nm (CB-P100H-LU01FA, insertion loss 2.2 dB, Proximion AB) and a negative dispersion of -9982 ps/nm (CB-N100H-LU01FA, insertion loss 3 dB, Proximion AB). The spectral phase modulation in the chirped pulse modulation was performed with an electro-optic phase modulator (EOSpace Inc., insertion loss 3.2 dB, electro-optic bandwidth 40 GHz), which was driven by the arbitrary waveform generator. The insertion loss of the chirped pulse modulation setup was 8.4 dB. \\

To spectrally filter the signal and idler photons, we used a programmable filter (Waveshaper 4000A, Finisar) to select an optical bandwidth of 10 GHz each, which introduced additional 4.5 dB of loss. The photons were detected using two superconducting nanowire single-photon detectors (EOS, Single Quantum, 17 ps timing jitter) connected to a time-to-digital converter (TimeTagger Ultra, Swabian Instruments, 18 ps timing jitter). An electronic reference signal (rectangular pulse with 10 ns duration, every 180 ns) was forwarded from the arbitrary waveform generator to the time-to-digital converter to assign photon arrival times to the segments of the radio-frequency waveform.

\subsection{Chirped pulse modulation}

Chirped pulse modulation (CPM) applies a dispersive Fourier transformation to temporally separate the spectral components of input pulses, enabling spectral phase modulation. The chirped pulse modulation affects both the temporal and spectral properties of the output pulses. Specifically, by applying a sinusoidal voltage \(V(t) = V_{0} \sin(\Omega t + \alpha)\) to the chirped pulse modulation, a time-bin at time \(t\) and frequency \(\nu\) is converted into coherent pulse copies of different order $m$ ($m = 0, \pm 1, \pm 2, \ldots $), which are shifted in time by \(m \cdot \Delta t\) and in frequency by \(m \cdot \Delta \nu\). Mathematically, this can be described by the transformation

\begin{equation}
    |\nu, t \rangle \rightarrow \sum_{m = -\infty}^{\infty} J_{m}(g) e^{-im\alpha} |\nu + m \Delta \nu, t + m \Delta t\rangle,
\end{equation}

where \(J_{m}(\circ)\) are the Bessel functions of the first kind and order \(m\) and \(g=V⁄V_{\pi}\) is the modulation depth. The time and frequency shifts are given by \(\Delta t = \beta_2 \Omega\) and \(\Delta \nu = \Omega/(2\pi)\), respectively, where \(\beta_2\) is the dispersion of the first CFBG. When two time-bins at the same frequency are launched into the chirped pulse modulation, temporally overlapping pulse copies at the output do not have perfect spectral overlap due to their spectral shift, diminishing the interference contrast. As shown in extended data Fig.~\ref{fig:Extended_Data_Figure_01}, this effect bounds the visibility to 95\% for 300 ps time-bin spacing (T-level qubits) and to 99\% for 100 ps spacing (t-level qubits). The phase \(\alpha\), which is controlled via relative timing adjustments of the RF waveform with respect to the time bins, enables precise control over the time-bin interference.

\pagebreak

\subsection{Tunable time-bin beam splitter operations}

An ideal time-bin beam splitter takes two input time-bins and maps them onto two output time-bins, each representing a coherent linear superposition of the two input time-bins. Chirped pulse modulation enables to shift time-bins both forward and backward in time — in turn implementing a time-bin beam splitter. By varying the modulation frequency, we implemented a time-bin beam splitter for the t-level qubits and T-level qubits. \\

With our chirped pulse modulation setup, we transformed time-bin encoded qubits (basis states \(|0\rangle, |1\rangle\)) into superposition states given by  \(J_0(g)|0\rangle + J_1(g) e^{i\alpha} |1\rangle\) and \(J_0(g)|1\rangle - J_1(g) e^{i\alpha} |0\rangle\) (up to normalization). By tuning the modulation depth $g$ and the phase \(\alpha\), we were able to implement a fully tunable time-bin beam splitter. To balance the superposition of two bins, we chose the first intersection point of the Bessel functions \(J_0\) and \(J_1\), which is given by \(g \approx 1.4342\). In our experiments, we used three beam splitter configurations: (i) measurement in the Z-basis (basis states \(|0\rangle, |1\rangle\))  by setting \(g=0\) and \(\alpha\) arbitrary, i.e. the time-bins remain unmodulated, (ii) measurement in the X-basis (basis states \(|+\rangle, |-\rangle\)) by setting \(g = 1.4342\) and \(\alpha = 0\) and (iii) measuring in the XY-plane (basis states \(|\alpha_{+}\rangle = (|0\rangle + e^{i\alpha}\ |1\rangle)/ \sqrt2, \ |\alpha_{-}\rangle = (|0\rangle-e^{i\alpha}|1\rangle)/ \sqrt2\)) by setting \(g = 1.4342\) and \(\alpha\) as the basis rotation angle. While the beam splitter settings (i) and (ii) are required to reconstruct the witness, the settings (iii) were used to certify one-way quantum operations. \\

In contrast to the ideal beam splitter, at the output of our system, photons were scattered into ancillary time-bins with different superposition states. The efficiency is independent of the phase and is given by $\eta(g) = |J_0(g)|^2 + |J_1(g)|^2$, yielding $\eta(0) = 1.0$ for measurement in the Z-Basis and $\eta(1.4342) = 0.601$ for measurements in the X-basis and the rotated basis. In the evaluation of our two-photon cluster state experiments, we used a correction factor to account for the scattering losses, i.e. each time we measured in the Z-basis, we multiplied the raw counts by a factor of 0.601.

\subsection{High-capacity quantum processing via spectral multiplexing}

Time-bin encoded qubits require a specific volume in the time-frequency domain, fundamentally constrained by their time-bandwidth product. In our encoding scheme, each qubit occupies approximately 25 GHz (corresponding to 0.2 nm) of spectral bandwidth and 200 ps of temporal bandwidth. Given that a chirped fiber Bragg grating with a dispersion of 10 ns/nm stretches the time-bins to 2 ns, the maximum repetition rate without introducing time-bin crosstalk is 500 MHz. Assuming an available optical bandwidth of ~5 THz (covering the full C-band), the resulting maximum processable information rate reaches 100 GigaQubits/s. 

\pagebreak

\subsection{Stabilizers and witness}

The stabilizer formalism provides a framework to quantify entanglement of multipartite quantum states~\cite{Toth_2005}, requiring fewer measurement settings compared with e.g. quantum state tomography~\cite{James_2001}. For our approach, we used the stabilizer formalism to construct a witness operator which diagnoses the presence of multipartite entanglement within the given quantum state. A negative expectation value of the witness operator certifies the presence of genuine multipartite entanglement~\cite{Horodecki_2001, Bourennane_2004}. For the 4-qubit cluster state that we consider here, a witness operator is given by~\cite{Kiesel_2005, Chen_2007}

\begin{equation}
    W_{C} = 2 - \frac{1}{2} \Big(11ZZ +ZZ11 + 1ZXX + Z1XX + XX1Z + XXZ1 \Big),
\end{equation}

where $X$ and $Z$ are Pauli operators and $1$ is the identity operator. The choice of this witness is not unique, however, it is particularly suitable to our experiment, as it exclusively requires measurement either on the t-level or on the T-level qubits and permits linear incoherent noise of up to 33.3\%. Moreover, this type of witness can be generalized to generic multi-partite and high-dimensional cluster states~\cite{Sciara_2019}, underlining the scalability of our approach. We estimated the statistical error of the stabilizers and witness using a Monte Carlo resampling method based on the measured counts. Assuming Poisson statistics, we generated $10^{6}$ samples to compute witness values for each sample and extracted the standard error from the resulting distribution. Moreover, based on the witness, a lower bound to the fidelity of our state with respect to the ideal cluster state is given by \(F \geq (1- W_{C})/2 = 90 \% \)~\cite{Toth_2005}.

\bmhead{Acknowledgements}

We acknowledge the correspondence with Adel Asseh and his colleagues from Proximion AB, who designed and manufactured the custom chirped fiber bragg gratings that were used in this work. 

\bmhead{Funding}

This research was funded by the European Union within the framework of the European Innovation Council’s Pathfinder program, under the project QuGANTIC, the European Research Council (ERC) within the project QFreC under the grant agreement no. 947603, and the German Research Foundation (Deutsche Forschungsgemeinschaft; DFG) within the cluster of excellence PhoenixD (EXC 2122, Project ID 390833453). This joint research project was financially supported by the State of Lower Saxony, Hannover, Germany.

\bmhead{Authors contributions}

P.R. and R.J. contributed equally. P.R. conceived the project in discussion with M.K. The experimental setup was designed and built by P.R. and R.J. with assistance form J.H. R.J. and P.R. performed the experiment, acquired and analyzed data. The theoretical analysis was performed by P.R. and O.V.M. The initial manuscript draft was written by P.R. and R.J. All authors contributed to the final version of the manuscript. The project was supervised by M.K.

\pagebreak
\section{Extended data figures}
\setcounter{figure}{0}
\renewcommand{\figurename}{Extended data Fig.}
\begin{figure}[h]
    \includegraphics[width=1.0\textwidth]{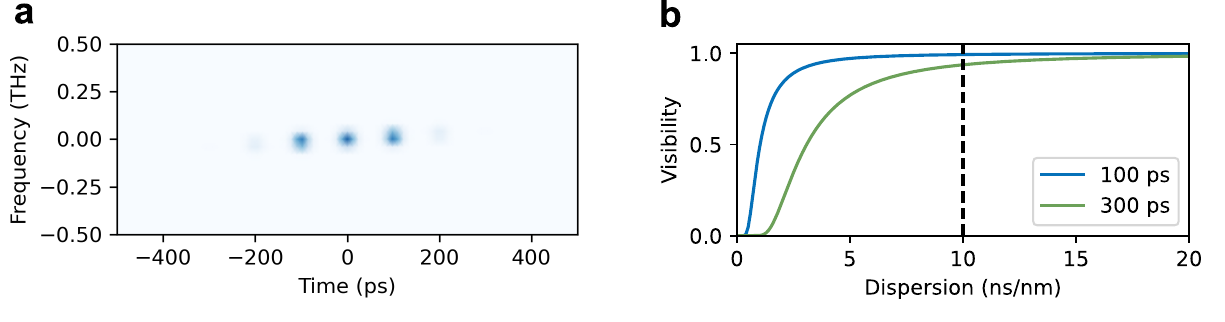}
    \centering
    \caption{\textbf{Spectral walk-off and visibility bounds in the chirped pulse modulation.} (a) Numerically calculated time-frequency spectrogram for a single Gaussian input pulse with 37 ps duration after being processed by the chirped pulse modulation (here, operated at a modulation frequency of 1.25 GHz and with 10 ns/nm dispersion). The center frequencies of the pulse copies at $\pm 100$ ps and $\pm 200$ ps are spectrally shifted with respect to the original input pulse at 0 ps. In the interference of two time-bin, this spectral walk-off bounds the interference contrast, i.e. the visibility. (b) Numerically obtained bound of the interference visibility for two time-bins as a function of the dispersion of the chirped fiber Bragg gratings. The curves correspond to the maximal attainable interference visibility of time-bins (each 37 ps full-width at half maximum) separated by 100 ps (blue, t-scale) and 300 ps (green, T-scale). The dashed black line indicates the dispersion value of the chirped fiber Bragg gratings used in this work ($\pm 10$ ns/nm).}
    \label{fig:Extended_Data_Figure_01}
\end{figure}

\begin{figure}[h]
    \includegraphics[width=1.0\textwidth]{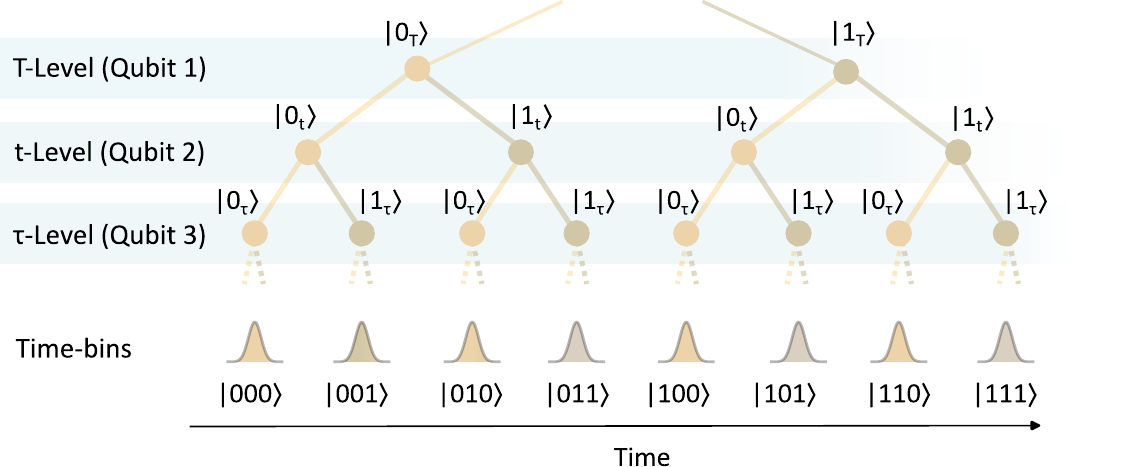}
    \centering
    \caption{\textbf{Scalability of the multi-level time-bin encoding scheme.} To encode more than one qubit into each photon, we leverage multi-level time-bin encoding. This encoding is represented by a binary tree, where each level of the tree encodes one qubit. On each level of the tree the left branch corresponds to basis state $|0 \rangle$ of the qubit, while the right branch corresponds to basis state $|1 \rangle$ of the qubit. In our case, to encode two qubits per photon, two levels (first level is called T-level, second level is called t-level) are required resulting in four time-bins. Further levels (such as the third $\tau$-level) can be introduced to encode even more complex states. To coherently manipulate the encoded qudits, a tunable time-bin beam splitter on the respective scale is required. In general, to encode $N$ qubits into one photon in the multi-level time-bin encoding, $2^{N}$ time-bins are required. The multi-level scheme can be extended to $d$-level qudits corresponding to $d$-ary trees.}
    \label{fig:Extended_Data_Figure_02}
\end{figure}

\begin{figure}[h]
    \includegraphics[width=1.0\textwidth]{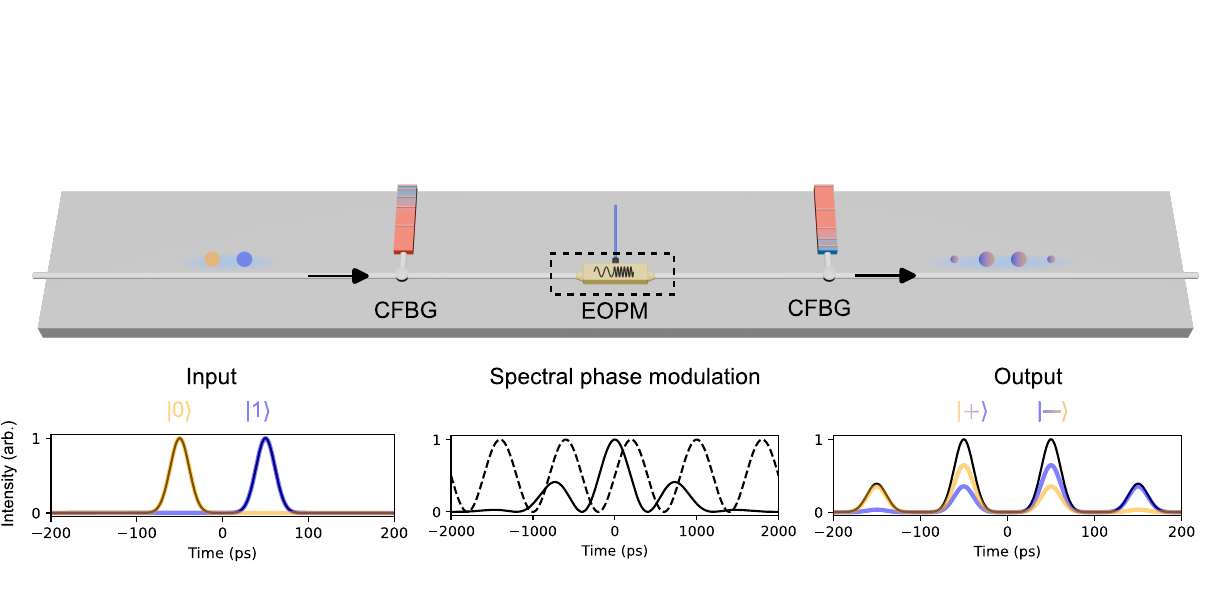}
    \centering
    \caption{\textbf{Interference of two time-bins via chirped pulse modulation.} Two input time-bins (left) with  100 ps separation (orange is the basis state $|0 \rangle$ and purple is the basis state $|1\rangle$) are dispersed with a chirped fiber Bragg grating (CFBG), implementing an approximate frequency-to-time mapping. On the dispersed pulse (center) the frequency components are mapped to different times and can spectral phase modulation is achieved by applying a radio-frequency signal (here, 1.25 GHz sinusoidal indicated by the dashed line) via the electro-optic phase modulator (EOPM). Recompression of the dispersed pulse with a CFBG of opposite dispersion, interferes the time-bins (right). The contributions of the initial time-bins to the superposition state (black solid line), i.e. the $|+ \rangle$ state in the left output time-bin and the $|- \rangle$ state in the right output time-bin, are indicated in orange and purple.}
    \label{fig:Extended_Data_Figure_03}
\end{figure}

\clearpage

\bibliography{bibliography}

\end{document}